\documentclass[useAMS,usenatbib,usegraphicx]{mn2e}
\usepackage{txfonts}
\usepackage{Journals}
\usepackage[usenames,dvipsnames]{color}

\pdfoutput=1
\def\eq#1{\begin{equation} #1 \end{equation}}

\def\E#1{\hbox{$10^{#1}$}}
\def\sub#1{_{\rm #1}}
\def\about  {\hbox{$\sim$}}
\def\x      {\hbox{$\times$}}
\def\mic   {\hbox{$\mu$m}}

\font\math = cmmi10
\def\m#1{\hbox{\math \char'#1}} 
\def\v{\m{166}}

\def\vK    {\hbox{$\v_{\rm K}$}}
\def\vKK   {\hbox{$\v^2_{\rm K}$}}
\def\vw    {\hbox{$\v_{\rm w}$}}
\def\vc    {\hbox{$\v_{\rm c}$}}
\def\Mw     {\hbox{$\dot M_{\rm w}$}}
\def\Mdot   {\hbox{$\dot M$}}
\def\Ncrit  {\hbox{$N_{\rm crit}$}}

\def\mH     {\hbox{$m_{\rm H}$}}
\def\Nmin   {\hbox{$N_{\rm H,min}$}}
\def\Mc     {\hbox{$M_{\rm c}$}}
\def\Rc     {\hbox{$R_{\rm c}$}}
\def\Ac     {\hbox{$A_{\rm c}$}}
\def\Nc     {\hbox{$N_{\rm H,c}$}}
\def\nc     {\hbox{$n_{\rm H,c}$}}
\def\Rc     {\hbox{$R_{\rm c}$}}
\def\Rd     {\hbox{$R_{\rm d}$}}
\def\LEdd   {\hbox{$L_{\rm Edd}$}}
\def\Lbol   {\hbox{$L_{\rm bol}$}}
\def\Mo    {\hbox{$M_{\odot}$}}

\def\cs     {\hbox{cm$^{-2}$}}
\def\kms    {\hbox{km\,s$^{-1}$}}
\def\erg    {\hbox{erg\,s$^{-1}$}}
\def\Ha    {\hbox{H$\alpha$}}
\def\Hb    {\hbox{H$\beta$}}
\def\LbHa  {\hbox{$L_{\rm bH\alpha}$}}
\def\LnHa  {\hbox{$L_{\rm nH\alpha}$}}
\def\LBL   {\hbox{$L_{\rm BL}$}}
\def\CBL   {\hbox{$C_{\rm BL}$}}
\def\Cw    {\hbox{$C_{\rm BL}^{\rm W}$}}
\def\Ck    {\hbox{$C_{\rm BL}^{\rm K}$}}
\def\Fw    {\hbox{$\phi^{\rm W}(\v)$}}
\def\Fk    {\hbox{$\phi^{\rm K}(\v)$}}

\title[BLR Evolution in AGNs]
            {Evolution of Broad-line Emission from
            Active Galactic Nuclei}

\author[Elitzur, Ho \& Trump]
       {Moshe Elitzur$^1$, Luis C. Ho$^{2,3}$ and Jonathan R. Trump$^{4,5,6}$\\
   $^1$Department of Physics and Astronomy,
       University of Kentucky,
       Lexington, KY 40506-0055,
       USA;
       moshe@pa.uky.edu\\
   $^2$Kavli Institute for Astronomy and Astrophysics,
       Peking University,
       Beijing 100871,
       China\\
   $^3$The Observatories of the Carnegie Institution for Science,
       813 Santa Barbara Street,
       Pasadena, CA 91101,
       USA;
       lho@obs.carnegiescience.edu\\
   $^4$University of California Observatories/Lick Observatory,
       University of California,
       Santa Cruz, CA 95064,
       USA\\
   $^5$Department of Astronomy and Astrophysics,
       Pennsylvania State University,
       University Park, PA 16802,
       USA;
       jtrump@psu.edu\\
   $^6$Hubble Fellow\\
       }

\date{Accepted 2013 December 16.  Received 2013 November 21; in original form 2013 October 1}
\pagerange{\pageref{firstpage}--\pageref{lastpage}} \pubyear{2013}

\begin{document}
\label{firstpage} \maketitle

\begin{abstract}
Apart from viewing-dependent obscuration, intrinsic broad-line emission from
active galactic nuclei (AGNs) follows an evolutionary sequence: Type $1 \to
1.2/1.5 \to 1.8/1.9 \to 2$ as the accretion rate onto the central black hole
is decreasing. This spectral evolution is controlled, at least in part, by
the parameter $L_{\rm bol}/M^{2/3}$, where $L_{\rm bol}$ is the AGN
bolometric luminosity and $M$ is the black hole mass. Both this dependence
and the double-peaked profiles that emerge along the sequence arise naturally
in the disk-wind scenario for the AGN broad-line region.
\end{abstract}

\section{Introduction}

Supermassive black holes are ubiquitous in present-day galaxies
\citep{Kormendy13}. Numerous observations find tight correlations between
properties of the host galaxy and its central black hole \citep{Ferrarese00,
Gebhardt00, Tremaine02, Marconi03, Greene06}, indicating that the two strongly
affect each other's structure and evolution. Understanding this link requires
studies of the evolution of supermassive black holes and their immediate
environments. Accretion onto the black hole produces observable nuclear
activity that may occur in all massive galaxies with a duty cycle of
\about\E{-2} \citep{Soltan82, Marconi04, Greene07}, and the AGN bolometric
luminosity $L_{\rm bol}$ is directly related to the accretion rate. Since
low-luminosity AGNs are not simply scaled-down versions of their more familiar
cousins \citep{Ho08}, the classical Seyfert galaxies and quasars, studying the
variation with $L_{\rm bol}$ of various AGN radiative signatures sheds light on
the evolution of the black hole environment as its accretion rate is
decreasing.

{Prominent signatures of the pc and sub-pc scales around the black hole include
spectral lines from the broad-line region (BLR) and infrared emission from the
dusty torus. Observations show that the two are just the inner and outer
regions, respectively, of a single, continuous distribution of clouds whose
composition undergoes a change at the dust sublimation radius
(\citealt{Elitzur08}, and references therein). Such a structure arises
naturally in the disk-wind scenario, which involves the outflow of clouds
embedded in a hydromagnetic disk wind \citep{Emmering92, Konigl94, Kartje96,
Bottorff97, Bottorff00, Everett05}. Outflows are a common component of AGNs
\citep[see][and references therein]{Everett07, Proga07} and a number of
individual sources provide direct observational evidence for the disk wind
structure. \cite{Hall03} find that only a disk wind outflow can explain the
segregation of broad absorption lines in the BAL QSO SDSS J0300+0048. From
detailed line profile variability studies, \cite{Kollatschny03} finds evidence
for an accretion disk wind in the BLR of the narrow line Seyfert 1 Mrk 110.
Analysis of line dispersion of variable broad line emission leads to a similar
conclusion also for NGC 5548 \citep{Kollatschny13}, the well-studied AGN whose
BLR was successfully modeled in great detail with a clumpy disk outflow by
\cite{Bottorff97}. Disk winds also provide satisfactory explanations for the
trends exhibited by high-ionization, broad emission lines from 30,000 SDSS
quasars \citep{Richards11} as well as the general properties of UV and X-ray
absorption lines in both  quasars and Seyfert galaxies \citep{Schurch09,
Fukumura10b}. All in all, magnetically driven winds seem to be the best
candidate for the origin of high-velocity winds emanating from the inner radii
of AGN accretion disks \citep{Peterson06, SloneNetzer12}.
}

An immediate consequence of the disk-wind scenario is the prediction that the
torus and the BLR disappear at low bolometric luminosities, i.e., low accretion
rates \citep{Elitzur_Shlosman}. The reason is that, as the mass accretion rate
decreases, the mass outflow rate of a disk wind with fixed radial column cannot
be sustained below a certain accretion limit. This conclusion, a direct
consequence of mass conservation, has been verified in observations. Torus
disappearance has been verified in low-luminosity AGNs by infrared observations
\citep{van-der-Wolk10, Trump11}, and the existence of ``true type 2''
AGNs---unobscured sources that lack broad-line emission---has been established
in studies by \cite{Tran01, Tran03}, \cite{Panessa02}, and \cite{Laor03}, among
others. In particular, \cite{ElitzurHo09} find that the BLR disappears when
$L_{\rm bol} \la 5\times\E{39}\,M_7^{2/3} \,\erg$, where $M_7 = M/10^7\Mo$.
Note that these ``true type 2'' AGNs differ from the ``obscured type 2'' class
in the historical AGN unified model \citep{Ski93}.  While all type 2 AGN have
similar optical spectra with narrow emission lines, the ``true type 2'' AGNs
intrinsically lack broad emission lines due to structural changes rather than
torus obscuration.  In this work we use ``type 2'' to refer to these unobscured
``true type 2'' AGNs.

\begin{figure*}
  \centering
  \includegraphics[width=\hsize,clip]{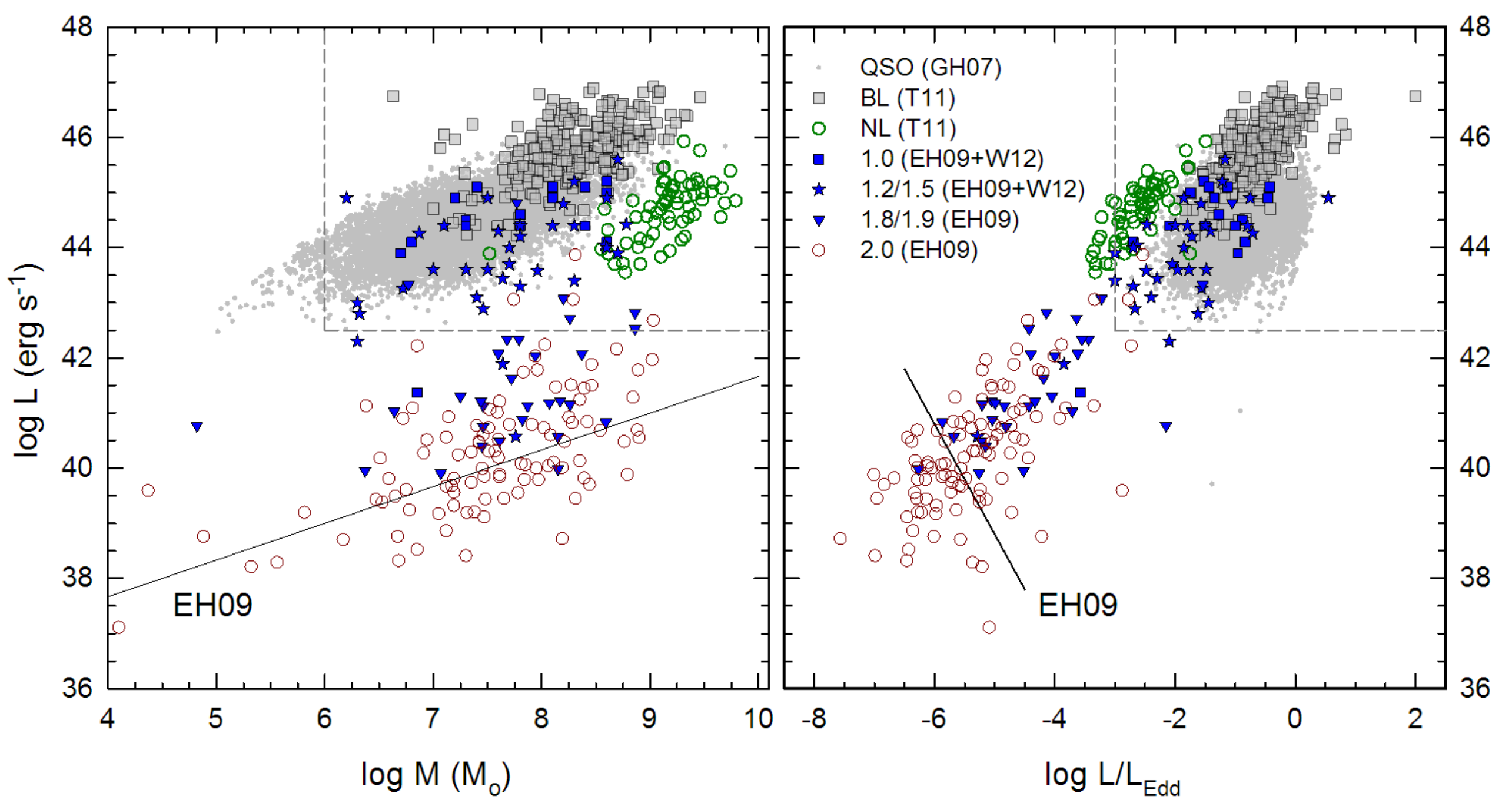}

\caption{Distribution of black hole masses $M$ ({\em left panels}) and
Eddington ratios $L_{\rm bol}/\LEdd$ ({\em right panels}) vs. bolometric
luminosity $L_{\rm bol}$ for objects separated by spectral classification.
Objects marked as ``QSO" refer to the sample of quasars and high-luminosity
Seyfert 1 nuclei studied by \citeauthor{Greene07} (2007; GH07). The ``BL" and
``NL" labels are for sources from the COSMOS survey classified as,
respectively, broad- and narrow-line emitters by \citeauthor{Trump11} (2011;
T11). Objects marked by the numerical classification of their AGN type are from
\citeauthor{ElitzurHo09} (2009; EH09) and \citeauthor{Winter12} (2012; W12).
The solid lines labeled EH09 show the bounds on broad-line emission found in
\citet{ElitzurHo09}: ({\em left}) $\log L = 35 + \frac23\log M$ and ({\em
right}) $\log L = 28.8 - 2\log (L/\LEdd)$. Dashed grey lines delineate the
domains of the \citet{Stern12a,Stern12b} sample, analyzed separately in
\S\ref{sec:SL}. \label{fig:L-M-Edd}
}
\end{figure*}

In a significant recent development, \cite{Stern12a, Stern12b} studied in
detail an SDSS-selected sample of type 1 AGNs and found that at low luminosity,
most of them actually appear as intermediate types. These observations show
that as the accretion rate decreases, the intrinsic AGN classification (apart
from torus obscuration) evolves from type 1 to an intermediate type, and
finally at low luminosity, potentially to pure type 2. Stern \& Laor note that
the observational evidence for true type 2 AGNs for some previously claimed
sources is not yet fully established on the basis of current limits on the
absence of broad \Ha\ emission. We agree with their assessment.  For example,
the upper limits to the broad H$\alpha$ luminosity for the majority of the type
2 sources in the Palomar survey \citep{Ho97a, Ho03} are not inconsistent with
the level of detected X-ray emission \citep{Ho09b}. But this just implies
ambiguity, not invalidation, since it does not mean that broad \Ha\ must be
there; the data are simply not conclusive in demonstrating the complete absence
of broad \Ha. Nevertheless, Stern \& Laor also noted that there are sources
where the absence of broad \Ha\ emission is established with such stringent
upper limits, well below the level expected from their X-rays, that they indeed
appear to be true type 2 AGNs. This is qualitatively consistent with the
detection of variable nuclear ultraviolet emission in type 2 LINERs
\citep{Maoz05} and X-ray cores with low levels of intrinsic absorption
\citep[][and references therein]{Ho08}.

Here we advance the view that changes in the accretion rate are responsible for
the observed relative strengths of broad emission lines in AGNs. We analyze all
the available data, including the Stern \& Laor sample, in \S2 and find
evidence for a consistent trend of intrinsic BLR evolution with luminosity. In
\S3 we show that the disk outflow scenario offers a natural explanation for the
observed spectral evolutionary sequence, and conclude in \S4 with summary and
discussion.

\section{Intermediate-type AGNs}
\label{sec:Intermediates}

Intermediate-type AGNs are objects that show broad-line emission at a lower
level relative to the narrow lines than in ordinary type 1 sources. The
intermediate classification 1.x was originally based on the flux ratio [{\sc
O\,iii}]$\lambda$5007/H$\beta_{\rm total}$, which increases from type 1.2 to
1.8; in type 1.9 AGNs \Ha\ is the only broad emission line seen, possibly
reflecting insufficient observational sensitivity \citep{Osterbrock81}.
High-resolution optical or ultraviolet images often reveal the presence of a
central point source in intermediates, although the detection difficulty varies
depending on the object, the amount of host galaxy dilution, and, importantly,
the image resolution \citep{Ho08}. The vast majority also contain X-ray cores
with little evidence for intrinsic absorption.

Historical unification posits that in type 1 AGNs we have a direct view of the
BLR and in type 2 the BLR is obscured \citep{Ski93}. A simple interpolation
might suggest that in intermediates the BLR is either partially obscured or
undergoes extinction by optically thin dust, but this contradicts the presence
of a visible central point source and the \cite{Trippe10} finding that the
majority of intermediates in their sample are inconsistent with either
reddening of the BLR or viewing along a line of sight that grazes the
atmosphere of a central dusty torus. \cite{Stern12b} find that at low
luminosity most type 1 AGNs appear as intermediate types, and this cannot be
explained by partial obscuration or extinction. These findings show instead
that as the accretion rate decreases, broad-line emission progressively
declines from its high intensity in quasars and Seyfert 1s to the lower
relative strength of intermediate-type AGNs, disappearing altogether at
sufficiently low luminosities.

\subsection{Data Analysis}
\label{sec:data}

To study this suggested evolution, in Figure \ref{fig:L-M-Edd} we plot the data
assembled in \citeauthor{ElitzurHo09} (2009; hereafter EH09) in the $L_{\rm
bol}$--$M$ plane and its equivalent in the $L_{\rm bol}$--$L_{\rm bol}/\LEdd$
plane, where \LEdd\ is the Eddington luminosity. We exclude from the sample
NGC~1068 and NGC~4388, which were formally classified as S1.8 and S1.9 by
\cite{Ho97a}, respectively, due to the presence of weak broad \Ha\ emission,
although this is scattered radiation from their well-known hidden type 1
nucleus \citep{Ski85, Young96c}. We return to this point below (see
\S\ref{sec:Observations}). Included are the 8,497 high-luminosity sources from
\cite{Greene07} marked as ``QSO." Except for the specific designation of
intermediate-type sources, the figure is a repeat of the one in EH09 where all
broad-line emitters were shown under the single banner of ``type 1." To
increase the statistical significance for intermediate types we have grouped
type 1.2 together with 1.5, and 1.8 with 1.9. For types 1--1.5 the figure
additionally includes 48 sources from \cite{Winter12}. This sample was selected
from the all-sky survey conducted with the {\em Swift} Burst Alert Telescope
(BAT) in hard X-rays and thus is largely unbiased toward both obscuration and
host galaxy properties. Finally, the figure includes also the sample of
unobscured AGNs drawn from the Cosmic Evolution Survey (COSMOS;
\citealt{Scoville07}) by \cite{Trump11}. Of these sources, the 82 marked as
``BL" show the broad-line emission typical of type 1 AGNs.  Meanwhile the 24
marked as ``NL" were presumed to be type 2 objects by Trump et al. (2011),
based on the absence of broad \Hb\ line in their spectra.  However, the
brighter and longer-wavelength \Ha\ line, needed for confirming this
assignment, falls outside the observed spectral range of the ``NL" sources,
such that they are just as likely to be intermediate-type AGNs misclassified
only by their limited spectral range.

\begin{figure}
  \centering
  \includegraphics[width=\hsize,clip]{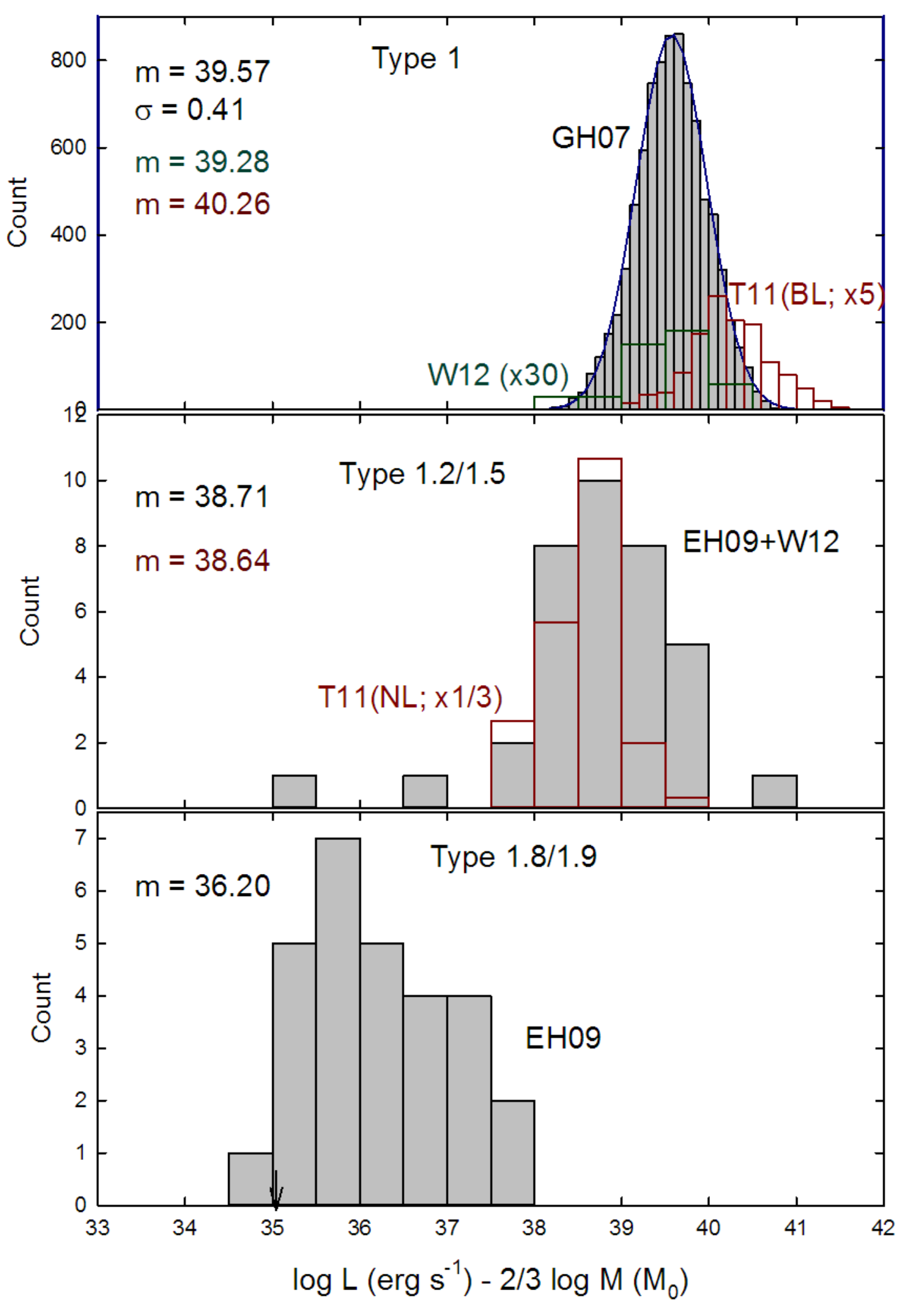}

\caption{Histograms of different classes of broad-line AGNs binned by $L_{\rm
bol}/M^{2/3}$. The distribution mean, $m$, is listed in each panel and the
arrow on the bottom axis marks the location of the EH09 bound shown in Figure
\ref{fig:L-M-Edd}. References for the data are listed in
Figure~\ref{fig:L-M-Edd}. The top panel shows the histogram of the type 1
sources from GH07 and lists its standard deviation $\sigma$ in addition to the
mean. A Gaussian plot with the listed values of $m$ and $\sigma$ is shown in a
blue line. The histograms of type 1 sources from W12 (multiplied by factor 30
for clarity and drawn in green) and the BL sources from T11 (multiplied by
factor 5, in red) are plotted separately. The T11 sample NL sources histogram
(in red, divided by 3) is shown in the middle panel together with the type
1.2/1.5 histogram from EH09 and W12. \label{fig:histograms}
}
\end{figure}

In spite of some possible mixing of type 1.x and type 2 AGNs, the figure
reveals an overall segregation between spectral classes of broad-line emitters.
Moreover, these classes appear to be stratified roughly in line with the EH09
boundary of broad-line emission, which is reproduced in the figure.  To test
the possible role of $L_{\rm bol}/M^{2/3}$ in AGN evolution, Figure
\ref{fig:histograms} shows the histograms of this variable for the different
broad-line spectral classes; histograms using the equivalent Eddington ratio
are redundant because they would involve the variable
$\Lbol\cdot(\Lbol/\LEdd)^2 \propto (\Lbol/M^{2/3})^3$. The QSO sample from
\cite{Greene07}, shown in the top panel, is sufficiently large to outline a
meaningful distribution, faithfully traced by a Gaussian  plotted in blue with
the sample mean and variance. The BL sources from \cite{Trump11} have a similar
distribution, drawn in red, although their histogram is slightly asymmetric
with a higher mean due to the higher luminosity limit of the parent
(high-redshift) survey. In contrast, the histogram of type 1 sources from
\cite{Winter12}, drawn in green, has essentially the same mean and, accounting
for the smaller sample size, shape as the QSO sample. Thus, the low-redshift
\cite{Winter12} compilation has a similar luminosity limit as the GH07 sample.
With the uncertainties surrounding their spectral classification, the NL
sources from \cite{Trump11} are shown in the middle panel together with the
combined sample of type 1.2/1.5 sources from W12 and EH09.  The two
distributions are similar, both in shape and mean $m$, giving support to the
conjecture that the NL sources might be actually intermediate-type rather than
type 2 AGNs.

Further statistical analysis for each spectral class is not warranted because
of the small sample sizes for all but the GH07 data set (notice the large
disparity between the histogram counts). Still, Figure \ref{fig:histograms}
establishes an unmistakable trend: When $L_{\rm bol}/M^{2/3}$ is decreasing,
AGN broad-line emission evolves away from type 1 toward higher values of the
intermediate sequence, disappearing altogether below the EH09 limit where all
AGNs become ``true type 2''.
\begin{figure*}
  \centering
  \includegraphics[width=\hsize,clip]{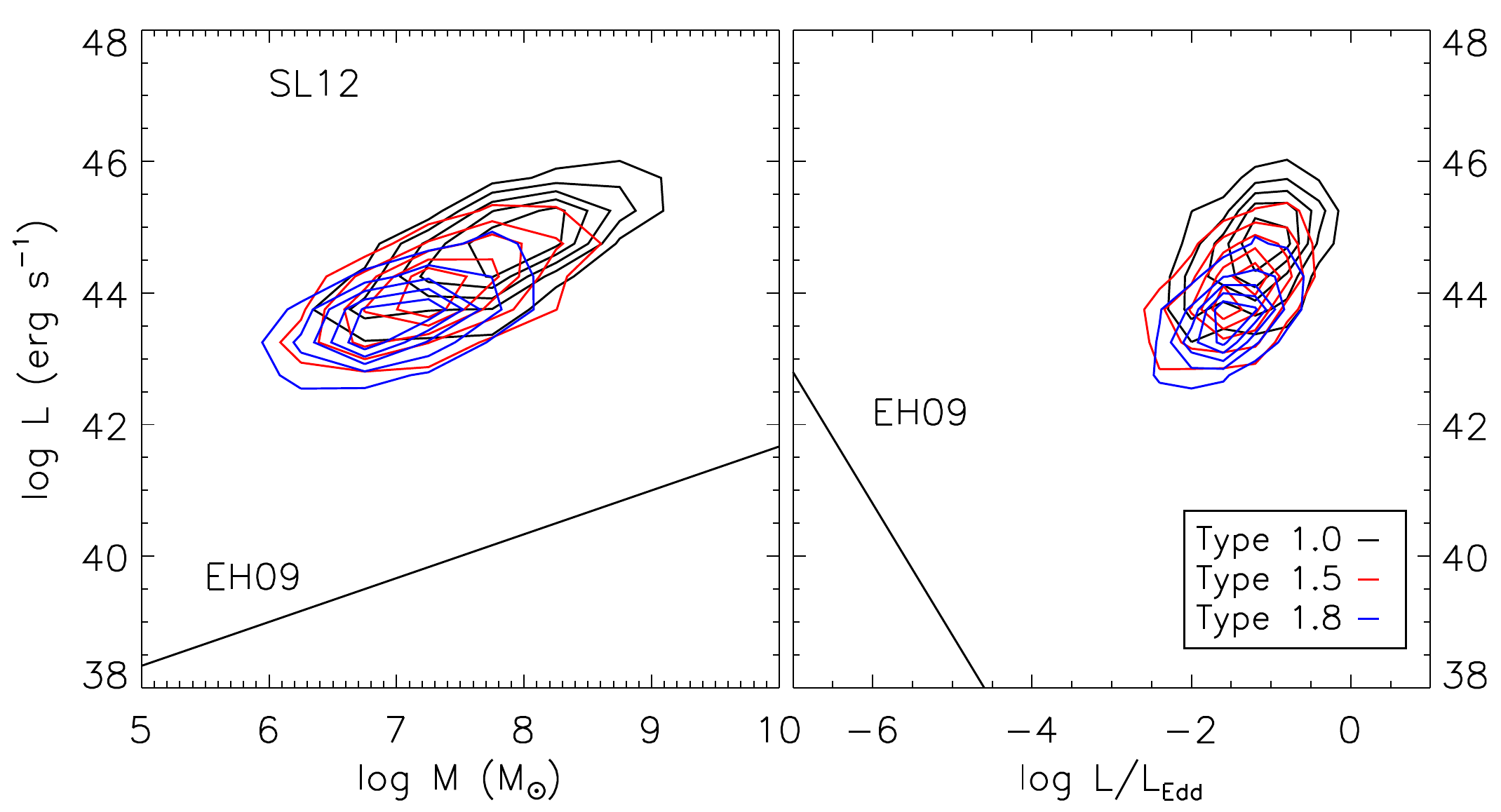}

\caption{Analog of figure \ref{fig:L-M-Edd} for the \citet{Stern12a,Stern12b}
sample (SL12). Because of the high density of points, the source distribution
of each AGN type is shown by colored contours evenly spaced in point density at
83, 67, 50, 33 and 17\% of the maximum. The EH09 boundaries are drawn for
guidance. Echoing Figure 1, there are significant changes in the mean AGN
properties from Type 1.0 to 1.5 to 1.8. \label{fig:SL_L-M-Edd}
}
\end{figure*}

\begin{figure}
  \centering
  \includegraphics[width=\hsize,clip]{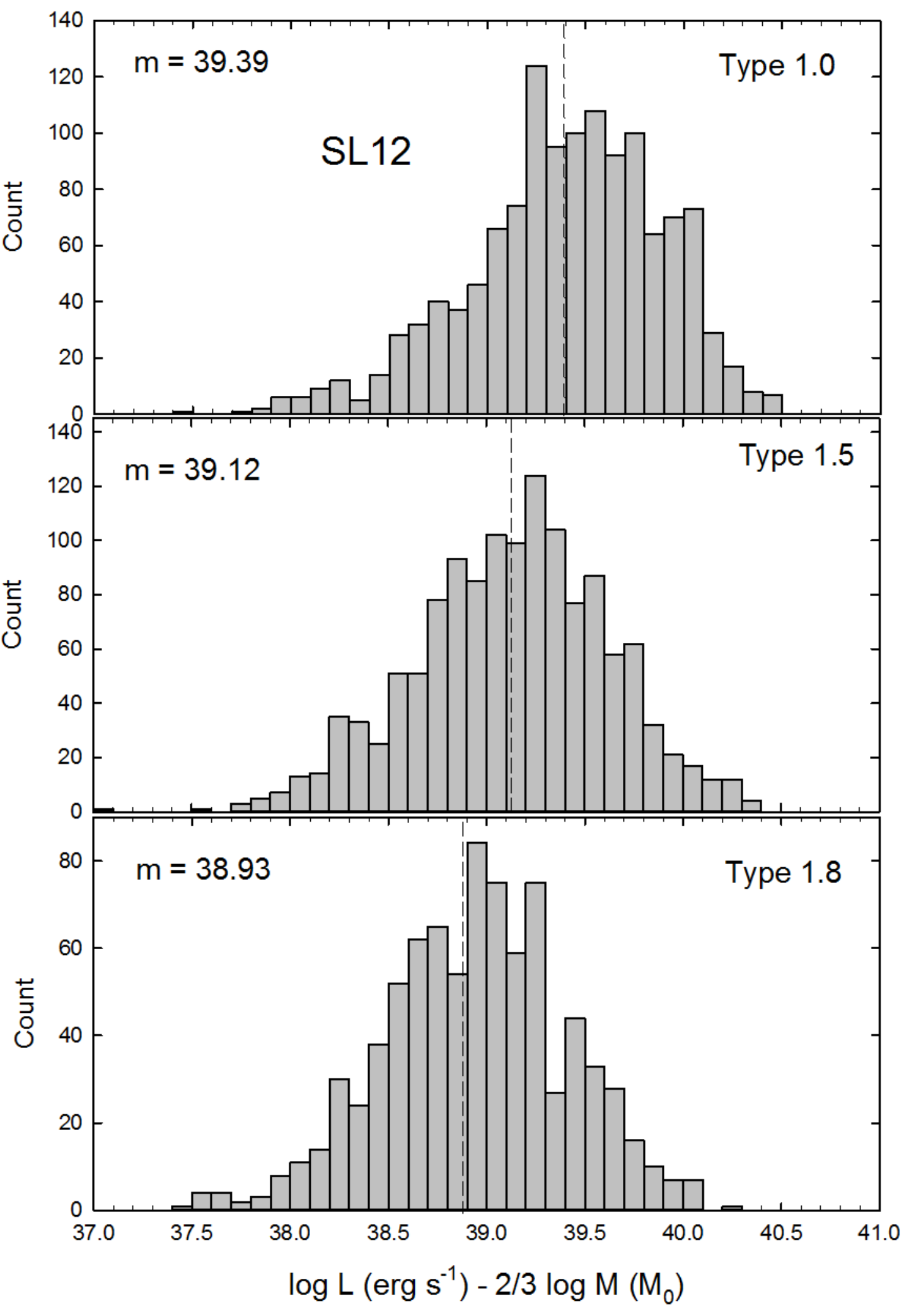}

\caption{Histograms for spectral classes of the SL12 sample, binned by $L_{\rm
bol}/M^{2/3}$. In each panel, the distribution mean $m$ is listed and drawn as
a vertical dashed line. \label{fig:SL_histograms}
}
\end{figure}

\subsection{The Stern \& Laor Sample}
\label{sec:SL}

\cite{Stern12a} have recently studied an SDSS-selected sample of 3,410 type 1
AGNs. Dashed grey lines in Figure \ref{fig:L-M-Edd} outline the regions of
parameter space where these sources fall. Although this sample (hereafter SL12)
does not extend to luminosities lower than \E{42} \erg\ and Eddington ratios
below \E{-3}, \cite{Stern12b} were still able to discern the transition to
intermediate spectral classes at decreasing luminosities. From a careful
decomposition of the \Ha\ profile into broad and narrow components, with
respective luminosities \LbHa\ and \LnHa, they show that the narrow component
starts dominating the total \Ha\ profile when the luminosity decreases, and
that at $\LbHa \la$ \E{42} \erg\ intermediate-type AGNs dominate the population
of broad-line emitters: At $\LbHa \ga$ \E{43} \erg\ the mean \Ha\ profiles are
typical of Seyfert 1.0s, at $\LbHa \sim$ \E{42} \erg\ they are typical of
Seyfert 1.5s, while at $\LbHa \sim$ \E{40}\,–-\E{41} \erg, broad-line objects
tend to be Seyfert 1.8s. Stern \& Laor also show that Seyfert 1.5s and 1.8s can
be defined analytically using the \LnHa/\LbHa\ ratio, with \LnHa/ \LbHa\
$\approx$ 0.1 marking the transition from type 1.0 to 1.5 and \LnHa\LbHa\
$\approx$ 0.3 the transition to 1.8.

Dr.\ J.\ Stern has kindly provided us the data for the full SL12 sample (Table
1 in \citealt{Stern12b}). We have classified the sources according to the Stern
\& Laor \LnHa/\LbHa\ criteria and found 1,266 type 1.0 objects, 1,306 type
1.5s, and 838 type 1.8s. Figure \ref{fig:SL_L-M-Edd} shows the $L_{\rm bol}-M$
and $L_{\rm bol}-L/\LEdd$ distributions for this sample, the analog of figure
\ref{fig:L-M-Edd}. Because of the high density of points, this sample is shown
with contour plots instead of a scatter diagram. Although limited to relatively
luminous objects, the SL12 sample still shows the same stratification by
spectral class as in Figure \ref{fig:L-M-Edd} and the same sequence in terms of
the variable $L_{\rm bol}/M^{2/3}$. These trends stand out more prominently in
Figure \ref{fig:SL_histograms}, which shows histograms binned in $L_{\rm
bol}/M^{2/3}$ for the three spectral classes, displaying a behavior similar to
that seen in Figure \ref{fig:histograms}. The mean for type 1.0 sources is
essentially the same as for the GH07 and W12 samples. For the intermediates,
the mean values are higher for the SL12 sample because the entire
low-luminosity end is missing. Still, the SL12 sample captures the trend of
decreasing $L_{\rm bol}/M^{2/3}$ along the spectral sequence
1.0$\to$1.5$\to$1.8 thanks to its large size. To assess the validity of the
differences between the mean values of $L_{\rm bol}/M^{2/3}$ we ran two
statistical tests for the null hypothesis that there are no inherent
differences between the objects in the three classes (i.e., all are drawn from
the same underlying population). The two-sample Kolmogorov-Smirnov (KS) test
for the 1.0 and 1.5 classes returned a $D$-statistic of 0.25, for a probability
$p = 9\x\E{-37}$ of being wrong in concluding that there is a true difference
in the two groups. The corresponding values for the comparison between the 1.5
and 1.8 classes were $D = 0.33$ and $p = 7\x\E{-62}$. Another widely used
method is the Mann-Whitney (MW; \citealt{Mann47}) rank sum test. For sample
sizes $n_1$ and $n_2$, the MW $U$-statistic returns values between 0 (no
overlap between the samples) and $0.5n_1n_2$. For the 1.0--1.5 comparison, the
MW test returned $U = 0.34n_1n_2$ for $p(\rm MW) = 1\x\E{-42}$, for the
1.5--1.8 comparison the results were $U = 0.39n_1n_2$ and $p(\rm MW) =
5\x\E{-19}$. Both tests show with overwhelming probabilities that the
differences between the mean values of $L_{\rm bol}/M^{2/3}$ for the three
spectral classes are real.

\begin{figure}
  \centering
  \includegraphics[width=\hsize,clip]{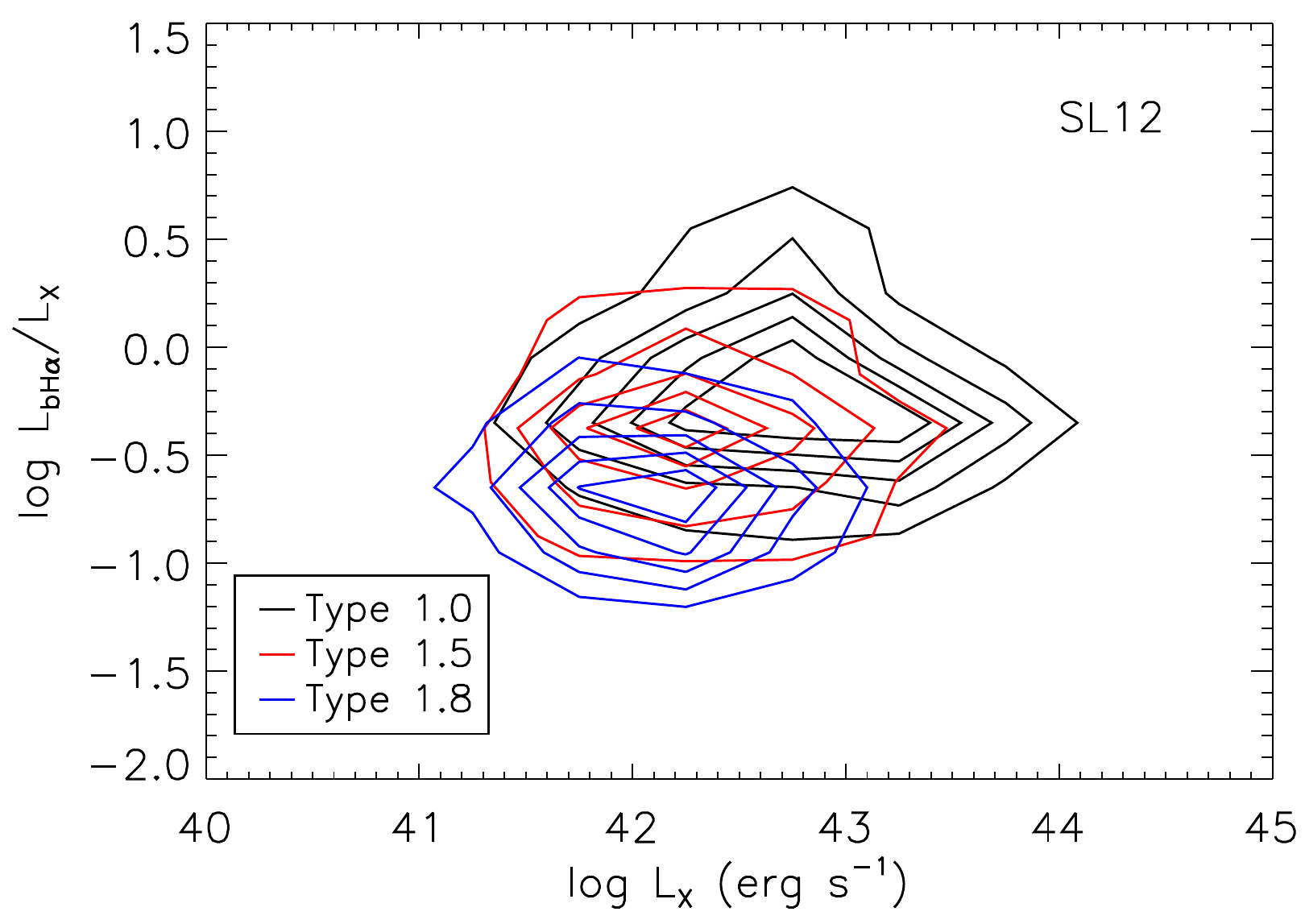}

\caption{Contours (as in Figure \ref{fig:SL_L-M-Edd}) for the SL12 sample of
the ratio of broad-line and 2 keV X-ray luminosities, equivalent to broad-line
covering factor, vs. X-ray luminosity, a proxy for the bolometric luminosity.
The mean broad-line ``covering fraction" decreases along the sequence from Type
1.0 to 1.5 to 1.8.
 \label{fig:SL_CF}
}
\end{figure}

\cite{Stern12b} suggest that the transition to intermediate types at decreasing
luminosity reflects an increase in the relative strength of the \Ha\ narrow
component rather than a decrease of the broad component emission. This
suggestion was based on their conclusion that \LbHa\ is proportional to the AGN
bolometric luminosity, i.e., the broad component ``covering factor"
$\LbHa/\Lbol$ is the same for all broad-line sources, independent of
luminosity. The evidence came from Figure 10 in \cite{Stern12a}, which displays
tight correlations between \LbHa\ and the AGN monochromatic luminosity $\nu
L_\nu$ in various spectral bands, from 2 keV all the way to 2.2 \mic. In
particular, the tightest proportionality is shown by the 2 keV X-rays (top-left
panel in Figure 10 of \citealt{Stern12a}), the spectral region least
contaminated by contributions from star formation. We reproduce this
correlation in Figure \ref{fig:SL_CF}, only instead of \LbHa\ we plot the ratio
$\LbHa/L_{\rm X}$, a proxy for the broad component covering factor, vs. $L_{\rm
X}$, which is generally regarded the most reliable indicator of AGN bolometric
luminosity. Importantly, instead of a scatter diagram with all sources marked
by the same symbol as done in \cite{Stern12a} we distinguish between the
spectral classes using different color contour plots. The figure shows a clear
trend of covering factor variation among the different spectral classes.
Because of the large scatter (more than 2 orders of magnitude) of the sample
data, lumping all the sources together in a single scatter diagram masked this
trend in the Stern \& Laor analysis.

\begin{figure*}
  \centering
  \includegraphics[width=0.7\hsize,clip]{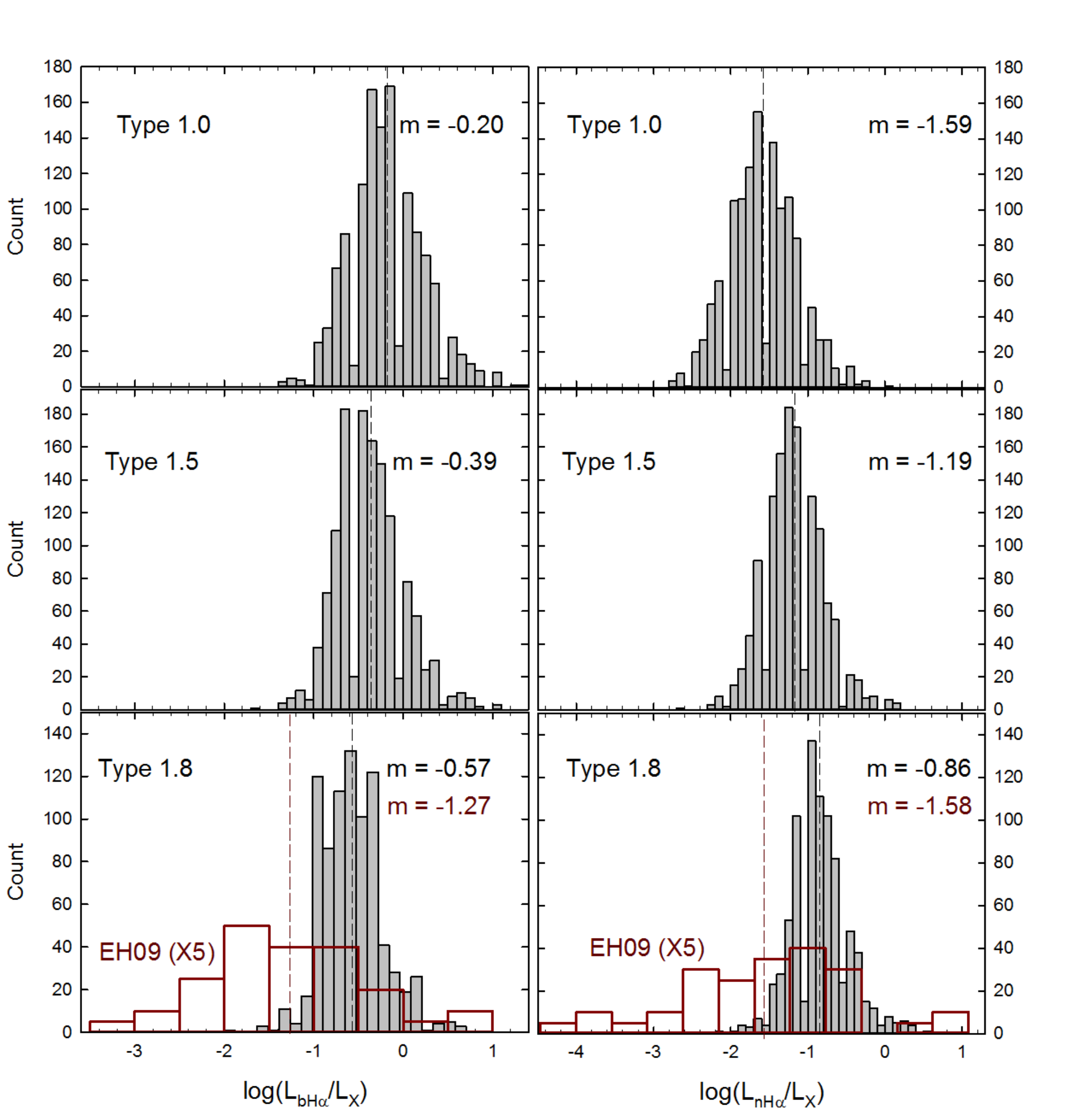}

\caption{``Covering factor" histograms of the \Ha\ broad ({\em left}) and
narrow ({\em right}) components for the SL12 sample. The spectral class is
marked in each panel and the mean value of the logarithm of ``covering factor"
is listed and drawn as a vertical dashed line. The lower panels show also (in
red) the histograms for all the intermediate-type sources in the EH09 sample;
see text for details. Note the great similarity between the mean values of
$\LnHa/L_{\rm X}$ of the EH09 sources and the SL12 type 1.0s.
\label{fig:SL_CF-histograms}
}
\end{figure*}

To tease out this effect more clearly, Figure \ref{fig:SL_CF-histograms} shows
histograms of the effective covering factors $\LbHa/L_{\rm X}$ and
$\LnHa/L_{\rm X}$ for the \Ha\ broad and narrow components, respectively, for
the SL12 sample. The right column of panels show that, as suggested by Stern \&
Laor, the narrow component covering factor indeed increases along the
1.0$\to$1.5$\to$1.8 sequence. This increase likely arises from the increased
prominence of the star formation contribution to \Ha\ emission relative to the
AGN when the latter's luminosity is decreasing.  As such, this effect reflects
external conditions, not any intrinsic variation in the AGN structure.
Meanwhile the left column of Figure \ref{fig:SL_CF-histograms} shows that,
together with the increase in the narrow component covering factor, the broad
component covering factor is {\em decreasing} along the 1.0$\to$1.5$\to$1.8
sequence. To assess the validity of this effect we again ran statistical tests
for the null hypothesis that there are no meaningful differences between the
broad \Ha\ effective covering factors of the three classes. For the 1.0--1.5
comparison the KS test returned $D(\rm KS) = 0.17$ for $p(\rm KS) =
1\x\E{-12}$, and the MW test produced $U(\rm MW) = 0.36n_1n_2$ and $p(\rm MW) =
1\x\E{-33}$. For the 1.5--1.8 comparison the corresponding results were $D(\rm
KS) = 0.23$ for $p(\rm KS) = 6\x\E{-20}$, and $U(\rm MW) = 0.35n_1n_2$ for
$p(\rm MW) = 4\x\E{-32}$. Both tests decisively show that the differences
between the classes are real. They are also highly significant: the difference
in $\log(\LbHa/L_X)$ of 0.37 dex implies that the broad component covering
factor of type 1.8 AGNs is, on average, less than half that in type 1.0.

We have attempted to verify that $\LbHa/L_x$ declines even further at lower
luminosities but only met with partial success for lack of sufficient data. The
EH09 sample contains 29 sources of type 1.8/1.9 and only 10 of type 1.2/1.5. An
additional 26 type 1.2/1.5 sources from the W12 sample have greatly aided  the
analysis in \S\ref{sec:data}, especially the histograms shown in Figure
\ref{fig:histograms}. Unfortunately, the W12 sample has not yet been subjected
to detailed spectral analysis to decompose the \Ha\ line to its broad and
narrow components, and thus could not be used for studying covering factors.
Faced with the inadequate number of 1.2/1.5 objects in the EH09 sample, we
decided to lump them together with the 1.8/1.9 sources for an overall set of 39
low-luminosity intermediate-type sources. All necessary data for these objects
are tabulated in \cite{Ho97a} and \cite{Ho09b}, including X-ray luminosity
integrated over 2--10 keV, which we converted to the Stern \& Laor 2-keV
$L_{\rm X}$ assuming an X-ray slope of $\Gamma = 1.9$ over this range.

The results are shown in Figure \ref{fig:SL_CF-histograms}, where the EH09
histograms are plotted in the bottom panels because the sample is dominated by
1.8/1.9 sources. The ratio $\LbHa/L_{\rm X}$ decreases further from its value
for the higher luminosity 1.8 AGNs of the SL12 sample, as expected from a
quantity reflecting intrinsic AGN variation. The decrease in mean values
between 1.8/1.9 sources in the SL12 and EH09 samples is about a factor of 70 in
luminosity, a factor of 440 in $L_{\rm bol}/M^{2/3}$ and a factor of 5 in the
effective broad-line covering factor $\LbHa/L_{\rm X}$. Because of the small
number statistics at the low-luminosity end this result cannot be taken too
literally, but the underlying effect is clearly established---the ratio
$\LbHa/L_{\rm X}$ decreases continuously and significantly with $L_{\rm
bol}/M^{2/3}$. In contrast, the variation of the ratio $\LnHa/L_{\rm X}$
between the two data sets is entirely out of line with the luminosity trend
within the SL12 sample. Remarkably, the average value of $\log \LnHa/L_{\rm X}$
for the EH09 sample ($-1.58$) is indistinguishable from that for the SL12 type
1.0 objects, whose narrow H$\alpha$ emission is presumably least complicated by
host galaxy contamination.  This reflects the fact that the Palomar spectra for
the EH09 objects---all nearby systems---properly isolate the emission from the
galaxy nuclei (Ho et al. 1997).  The $\LnHa/L_{\rm X}$ values for the EH09
sample should not be significantly contaminated by host galaxy emission.

The systematic change of $L_{\rm bH\alpha}/L_{\rm X}$ with AGN type seen in the
left panel of Figure 6 is not due to variations of $L_{\rm bol}/L_{\rm X}$ with
luminosity.  While the X-ray bolometric correction does change with decreasing
luminosity, the variation is too small to account for the observed trend. Over
the entire 2--10 keV luminosity range of \E{42}--\E{45} \erg\ covered by the
Stern \& Laor sample, $L_{\rm bol} \approx 45 L_{\rm 2-10\,keV}$ within $\pm
5\%$ \citep{Runnoe12}, while the mean value of the ratio $L_{\rm
bH\alpha}/L_{\rm X}$ decreases by more than a factor of 2 between type 1.0 and
1.8 sources. And for the extreme luminosities of the EH09 sample, which span
$L_{\rm 2-10\,keV} \approx$ \E{38}--\E{42} \erg, the X-ray bolometric
correction changes only by a factor of \about 3, to $L_{\rm bol} \approx 16
L_{\rm 2-10\,keV}$ (Ho 2008), yet the mean value of $L_{\rm bH\alpha}/L_{\rm
X}$ is down by a factor of 12 from that of the 1.0 sources.

\begin{figure}
  \centering
  \includegraphics[width=\hsize,clip]{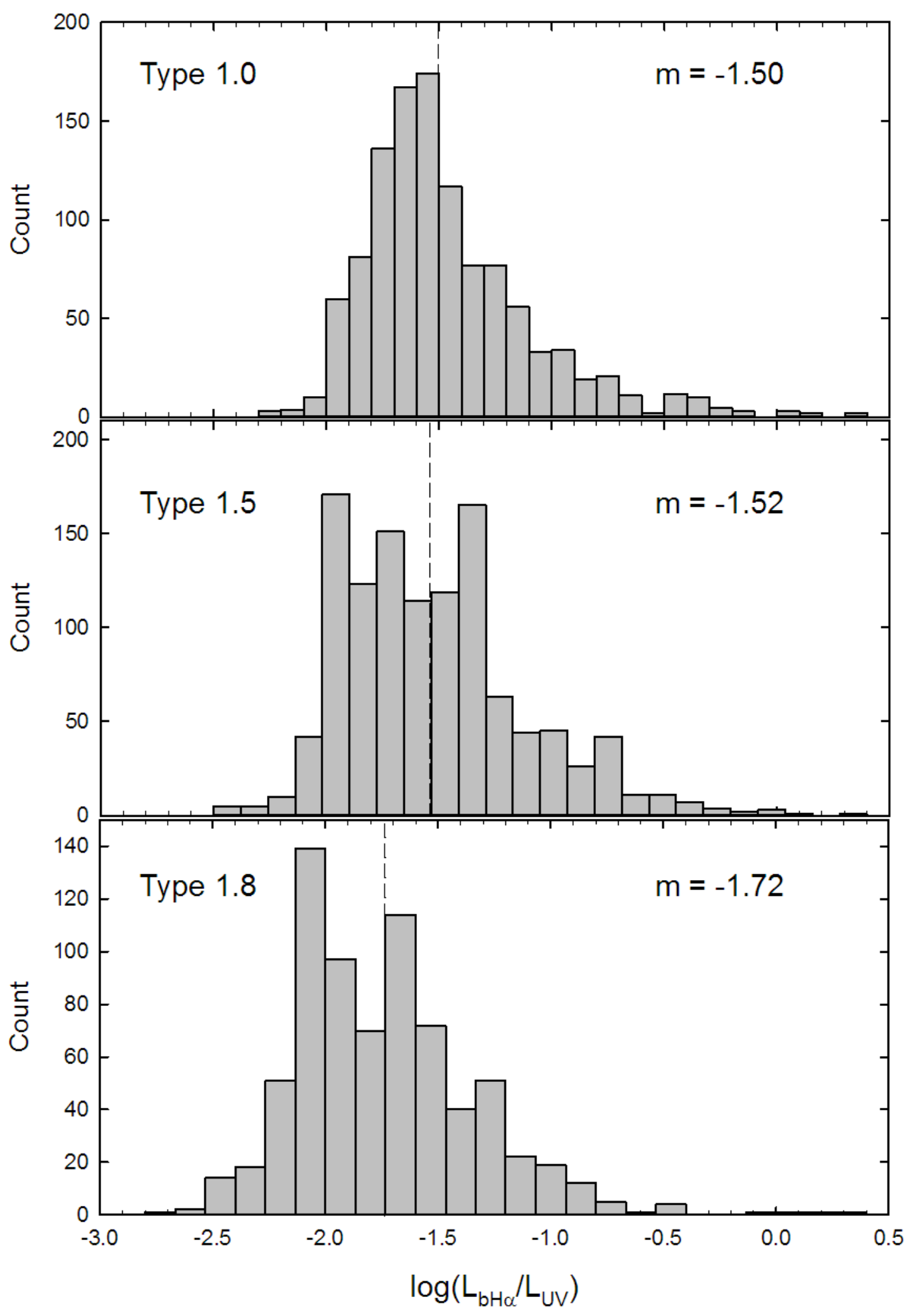}

\caption{``Covering factor" histograms for the SL12 sample broad  \Ha\
component similar to figure \ref{fig:SL_CF-histograms}, only for UV instead of
X-ray luminosity. \label{fig:SL_bHa-UV}
}
\end{figure}

As an additional check of the veracity of this result we have repeated the
analysis substituting UV (1528\AA\ monochromatic luminosity) for the X-rays.
Not all of the SL12 sources have UV data, reducing slightly the sample sizes
for 1.0, 1.5 and 1.8 classes to 1119, 1165 and 736, respectively. The results
are shown in Figure \ref{fig:SL_bHa-UV}, presenting the same data as Figure 7
of \cite{Stern12b} but in a separate histogram for each class instead of a
single scatter diagram for all sources. The similarity with the trend seen in
Figure \ref{fig:SL_CF-histograms} is evident and the statistical tests confirm
what is discernible to the eye. Comparing the 1.0 and 1.5 classes, the KS test
returns $D(\rm KS) = 0.10$ for a probability $p(\rm KS) = 2\x\E{-5}$ of being
wrong in concluding that the two are different. On the other hand, the MW test
produces $U(\rm MW) = 0.49n_1n_2$ and $p(\rm MW) = 0.47$; this test cannot
exclude the possibility that the difference between the two sets is due to
random sampling variability. The Mann Whitney test is inconclusive because it
is mostly sensitive to differences between the sample medians, which are the
same ($-1.60$) for both. In contrast, the Kolmogorov Smirnov test is decisive
because it compares the distribution shapes, and these are significantly
different from each other. There are no such ambiguities in the comparison of
the 1.5 and 1.8 samples, which yields $D(\rm KS) = 0.22$ for $p(\rm KS) =
2\x\E{-12}$ and $U(\rm MW) = 0.35n_1n_2$ for an overwhelming $p(\rm MW) =
2\x\E{-110}$ (the returned $U$-value is 22$\sigma$ away from the mean); these
are very clearly different types of objects.

Thus the UV data reaffirm the trend established through the X-rays. Denote by
$L_m$ the monochromatic luminosity in either X-ray or UV band. Then $\LbHa/L_m
= \zeta_m\LbHa/\Lbol$, where $\zeta_m = \Lbol/L_m$ is the bolometric correction
for the corresponding band. Going from high to low luminosities, the bolometric
corrections display the opposite behavior in the two bands --- $\zeta_m$ is
decreasing for X-rays and increasing for UV. Yet both bands display the same
trend of variation of $\LbHa/L_m$ with decreasing luminosity over the SL12
sample, showing conclusively that \LbHa/\Lbol\ is decreasing through the
spectral class sequence 1.0$\to$1.5$\to$1.8. A potential slight increase in
$\zeta_{\rm UV}$ throughout the SL12 sample could partly offset the intrinsic
decrease of \LbHa/\Lbol, bringing the mean values for the 1.0 and 1.5 classes
into equality and making the MW test inconclusive, although the KS test remains
decisive. Such an increase would still be insufficient to bridge the larger gap
to the 1.8 sources, where both tests remain overwhelmingly decisive.

\begin{figure}
  \centering
  \includegraphics[width=\hsize,clip]{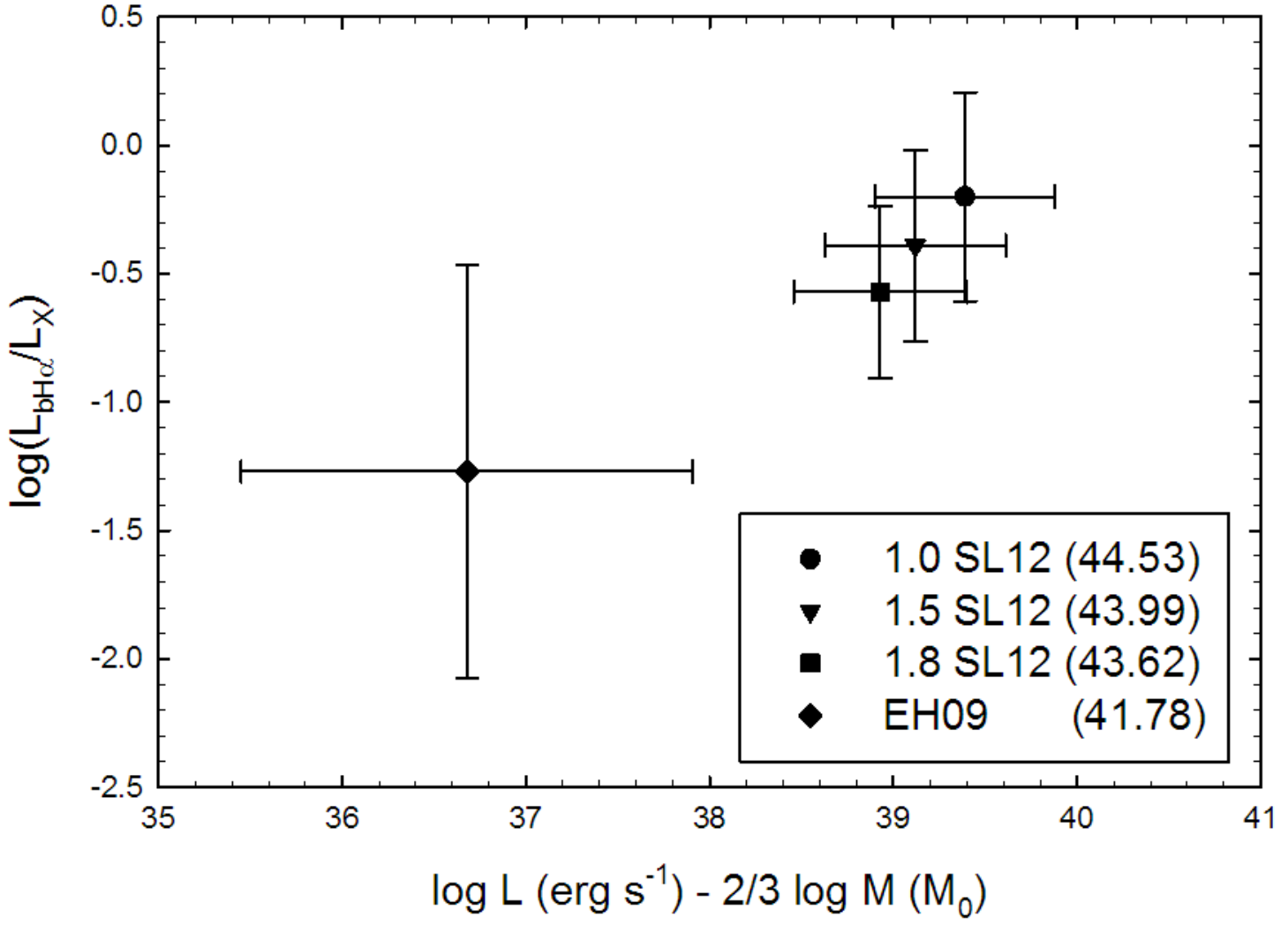}

\caption{Scatter diagram for the mean of broad \Ha\ ``covering factor" for the
various families of broad-line sources identified by their labels. Error bars
indicate one standard deviation. The numbers in parentheses are the means of
$\log \Lbol(\erg)$ for each group. \label{fig:CFplot}
}
\end{figure}

There is no tabulation of UV data for the EH09 sources, so we maintain
$\LbHa/L_{\rm X}$ as our primary indicator for the conversion efficiency of
continuum radiation to broad line emission. Figure \ref{fig:CFplot} summarizes
the results. The data establish, with high significance, that the average
effective covering factor $\LbHa/L_{\rm X}$ decreases  with $L_{\rm
bol}/M^{2/3}$ along the 1.0$\to$1.5$\to$1.8 sequence, albeit with scatter
causing some overlap among the AGN types. As is evident from Figure
\ref{fig:L-M-Edd}, the SL12 data set samples only the highest luminosity end of
intermediate-type AGNs and thus captures only the high-end boundaries of the
1.5 and 1.8 luminosity distributions. But thanks to its sizeable number of
sources, the differences it shows between those boundaries are highly
significant as verified by independent statistical tests. Among the SL12
sources, the covering factor decreases by more than a factor of 2 from type 1.0
to 1.8. The EH09 sample shows that the decrease continues further at lower
luminosities, its mean value of $\LbHa/L_{\rm X}$ is a factor of 12 lower than
for the type 1.0 sources. Even accounting for a decrease in the X-ray
bolometric correction and for the fact that the small size of the EH09 sample
implies that its mean value is less well-constrained, it seems safe to conclude
that, on average, $\LbHa/L_{\rm X}$ is at least a factor of 3 smaller for type
1.8/1.9 AGNs than for type 1.0. Since the immediate vicinity of the central
black hole is the only likely source of high-ionization lines with widths in
excess of 1,000 km s$^{-1}$, the decline of the broad \Ha\ covering factor
reflects a \emph{significant intrinsic change in the BLR at decreasing
luminosities.}

\subsection{Double-Peaked Emission in Intermediate Types}

As shown in the previous section, low-luminosity, low-Eddington ratio systems
with broad emission lines are typically of the type 1.8 or 1.9 variety. Another
notable property of intermediate types involves double-peaked broad-line
profiles, attributed to emission from rotating disks. Such profiles are found
in only about 3\% of the general broad-line AGN population \citep{Strateva03}
and most double-peak emitters have accretion rates considerably lower than the
Eddington limit \citep{Eracleous03, Eracleous09}. Conversely, low-luminosity,
low-Eddington ratio systems have a fairly high detection rate of double-peaked
profiles \citep{Ho00, Ho08}. Significantly, many of the 1.8 and 1.9 AGNs show
double-peaked broad-line profiles while most types 1.2/1.5 do not.  The
detection of double-peaked broad H$\alpha$ in low-luminosity AGNs is greatly
facilitated by small-aperture spectroscopy afforded by the {\it Hubble Space
Telescope (HST)}\ \citep[e.g.,][]{Ho00, Shields00}. Unfortunately, a uniform
{\it HST}\ spectroscopic census of the EH09 sample is not available, and thus
we do not have robust statistics on the incidence of double-peaked broad
H$\alpha$ emission in these objects.

\section{A Two-Component BLR}

{General considerations of AGN broad line modeling suggest that the emission
originates from a collection of clouds \citep[e.g.,][]{Netzer90}. Significant
contributions from the intercloud medium are precluded by its lower density,
which implies a higher ionization parameter. Direct observational evidence for
the clumpy nature of the BLR comes from studies of X-ray variability
\citep{Risaliti02} and from a detailed study of a set of absorption lines in a
quasar outflow \citep{Hall07}.} As shown above, two phenomena stand out along
the 1.0 $\to$ 1.2/1.5 $\to$ 1.8/1.9 sequence: the fraction of bolometric
luminosity converted to \Ha\ broad-line emission decreases, and double-peaked
profiles, signatures of rotating disk emission, appear. The simultaneous
emergence of two such disparate effects can be understood if as the luminosity
is decreasing, the BLR cloud distribution is gradually losing the components
with the highest elevations above the disk. Cloud motions near the disk surface
can be expected to reflect more closely Keplerian velocities, which is why
double peaked profiles would emerge in that case. Additionally, lower
elevations also reduce the line emission strength because the BLR geometrical
covering factor, the fraction of the AGN sky covered by clouds, becomes
smaller, intercepting a smaller fraction of the ionizing continuum. On top of
that, the ultraviolet/optical continuum emission from the AGN accretion disk
decreases towards its equatorial plane \citep{Laor89, Sun89, Kawaguchi10}, an
effect that was demonstrated convincingly by \cite{Risaliti11} in a recent
study of [{\sc O\,iii}] equivalent widths. The combined effect of a smaller
geometrical covering and a decrease in the ionizing continuum intensity
enhances the impact that a reduction in the height of the cloud distribution
has on broad-line emission.

\cite{Kartje99} were the first to note that such a structural change in the
cloud distribution occurs naturally in the disk-wind scenario at lower
luminosities. We now describe this effect.

\subsection{Cloud Trajectories}
\label{sec:trajectories}

{Disk outflow models are based on the seminal \cite{Blandford_Payne}
self-similar solution for a cold hydromagnetic wind driven centrifugally along
rotating magnetic field lines anchored in the disk. \cite{Rees87} pointed out
that if such winds were seeded with un-magnetized clouds, a configuration known
as ``melon seed'' plasma, they could offer a solution to the long standing
problem of BLR cloud confinement. In that case the clouds move as diamagnetic
blobs that can exclude (``push aside'') the field while being confined by the
magnetic pressure of the surrounding plasma. Adopting this approach,
\cite{Emmering92} extended the Blandford \& Payne solution to clumpy disk
outflows. Accounting for the change in composition across the dust sublimation
radius
\eq{\label{eq:Rd}
   \Rd \simeq 0.4 L_{45}^{1/2}\ \hbox{pc},
}
where $L_{45} = \Lbol/\E{45}\,\erg$ \citep{AGN2}, the BLR and torus were
unified in a single outflow: At radii beyond \Rd\ the clouds are dusty and
comprise the AGN obscuring torus, while the wind regions between \Rd\ and some
$R\sub{in} < \Rd$ contain the broad-line emitting clouds. Emmering et al showed
that modeling the BLR as a clumpy disk outflow yields good agreement with
observations (see \S\ref{sec:profiles} below).

\cite{Kartje99} considered similarly a wind that uplifts by its ram pressure
and confines by its magnetic pressure dense clouds fragmented from the disk.
Since the focus of their work was on water maser emission, Kartje et al
considered clouds at $r > \Rd$ while noting that they differ from BLR clouds
only in their composition, which does not affect the dynamics. But whereas
Emmering et al assumed the clouds to move along the outflow streamlines, Kartje
et al dropped this assumption and studied the actual motions of diamagnetic
blobs carried by the wind. Even if there were no field in the clouds
originally, it would gradually penetrate and break up the clouds, but this
process is expected to be slower than the sound speed within the clouds
\citep{Rees87}. Following Kartje et al we ignore this effect and consider the
motion of ``melon seed'' clouds confined by the wind magnetic pressure while
accelerated by its ram pressure against the gravitational pull of the central
black hole. The equation of motion of such a cloud with mass \Mc, velocity
vector ${\bf v_{\rm c}}$ and cross-sectional area \Ac\ normal to its motion
relative to the wind is then
\eq{\label{eq:EOM1}
   \frac{d{\bf v_{\rm c}}}{dt} \approx -\frac{GM}{r^2}{\bf \hat{r}} +
   \frac{\rho_{\rm w}\Ac}{M_{\rm c}}
   \left| {\bf v_{\rm w}} - {\bf v_{\rm c}} \right|
   ({\bf v_{\rm w}} - {\bf v_{\rm c}}) \, ,
}
where ${\bf \hat{r}}$ is unit vector in the radial direction, and ${\bf v}_{\rm
w}$ and $\rho\sub{w}$ are the wind velocity and mass density, respectively
\citep{Kartje99}. The cloud properties enter only through the column density
\hbox{$\Nc = \Mc/(\mH\Ac)$}, where \mH\ is the proton mass, that characterizes
the cloud whatever its geometrical shape. For spherical uniform clouds, \Nc\ is
$\frac23$ the column along the diameter. In the case of a thin, extended slab,
averaging over orientations yields \Nc\ that is twice the column along the slab
normal.

The wind density $\rho_{\rm w}$ can be found from its mass outflow rate \Mw\
through mass conservation. For the \citeauthor{Blandford_Payne} solution, $\Mw
= 4\pi r^2\rho\sub{w}\vw\ln(\Rd/R\sub{in})$. Since the BLR relative thickness
is likely $\Rd/R\sub{in} \la 100$ \citep{Emmering92}, the logarithmic factor is
probably $\la 4$ and is omitted for simplicity. The only property of the clouds
that enters is their column density. This suggests that we convert also \Mw\ to
an equivalent column density. The ratio $\Mw/(r\vK)$, where $\vK =
(GM/r)^{1/2}$ is the local Keplerian velocity, has dimensions of surface
density; thus we introduce
\eq{\label{eq:Ncrit1}
   \Ncrit = {\Mw\over4\pi\mH\, r\vK}
          = \Ncrit(\Rd)\left({\Rd\over r}\right)^{1/2},
}
where the factor $4\pi$ is added for convenience. The cloud equation of motion
(eq.\ \ref{eq:EOM1}) can then be written as
\eq{\label{eq:EOM2}
   \frac{d{\bf v_{\rm c}}}{dt} \approx -\frac{\vKK}{r}{\bf \hat{r}} +
   {\vKK\over r}{\Ncrit\over\Nc} {1\over\vK\vw}\left | {\bf v_{\rm w}} -
   {\bf v_{\rm c}} \right | ({\bf v_{\rm w}} - {\bf v_{\rm c}}) \, .
}
Clouds carried along the wind streamlines have ${\bf v_{\rm c}} \| {\bf v_{\rm
w}}$. Projecting on the wind direction, the equation of motion for such clouds
becomes
\eq{\label{eq:EOM3}
    \frac{d\vc}{dt} \approx \frac{\vKK}{r}
    \left[{\Ncrit\over\Nc} {\vw\over\vK}\left(1 - {\v_{\rm c}\over\vw}\right)^2
          - \cos\theta
    \right],
}
where $\theta$ is the angle between the wind velocity and the radial direction.
For a cloud to be accelerated along a streamline by the wind ram pressure, the
quantity inside the square brackets must be positive.
}

\begin{figure}
  \centering
  \includegraphics[width=0.9\hsize,clip]{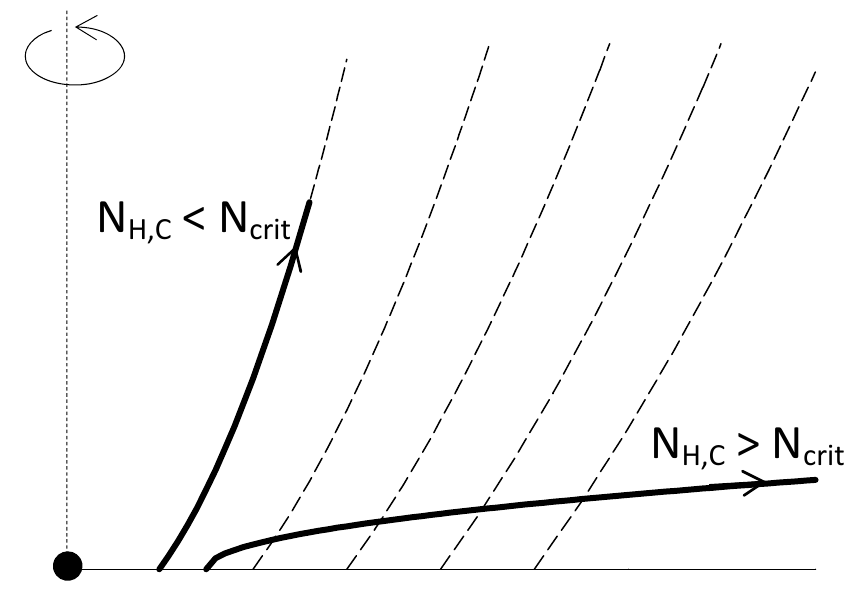}

\caption{Schematic representation of clumpy outflow from an AGN accretion disk.
Dashed lines show wind streamlines projected into the poloidal plane.
Trajectories of clouds uplifted by the wind ram pressure, shown by solid heavy
lines, depend on their column density \Nc. Subcritical clouds, with $\Nc <
\Ncrit$ (eq.\ \ref{eq:Ncrit1}), tend to follow the wind streamlines while
supercritical clouds, $\Nc > \Ncrit$, tend to move across streamlines and
remain close to the disk surface, as first noted by Kartje et al (1999; see
their Fig.\ 7). \label{fig:cartoon}
}
\end{figure}

Consider first the cloud initial motion near the disk surface, then $\theta$ is
the wind launch angle measured from the disk surface (typically
$\about\,20\degr$ in the Blandford \& Payne solution). The magnitude of the
wind launch velocity is comparable to the local turbulence velocity in the
disk, which is roughly 10\% of the local Keplerian velocity
\citep{Elitzur_Shlosman}; that is, $\vw \ll \vK$ close to the disk (in the
Blandford \& Payne solution, the launch velocity of the cold wind is taken as
0). Therefore, the term in square brackets in equation \ref{eq:EOM3} can be
positive only when $\Nc < \Ncrit$; clouds must be subcritical to accelerate
along the wind streamlines. To ensure that such clouds can follow the rising
trajectory shown in Figure \ref{fig:cartoon} we must verify that their
acceleration remains positive away from the disk. Indeed, both relevant ratios
increase as the cloud rises. The ratio \vw/\vK\ increases because of the wind
acceleration. Confinement by the wind magnetic pressure implies that the gas
density inside clouds obeys $\nc \propto B^2$. With a constant cloud mass and
the outflow magnetic field $B\propto r^{-5/4}$ from Blandford \& Payne, the
radial variation of the cloud properties is $\nc \propto r^{-5/2}, \Rc \propto
r^{5/6}$ and $\Nc \propto r^{-5/3}$. Therefore the ratio \Ncrit/\Nc\ increases
with distance as $r^{7/6}$ --- clouds that start subcritical will remain so. It
is also important to note that, in spite of the magnetic pressure confinement,
the cloud column density decreases as it rises above the disk. The vertical
range of broad-line emission is thus limited, resulting in a toroidal geometry
for the BLR.

{Supercritical clouds, those with $\Nc > \Ncrit$, cannot float with the wind.
Instead, they tend to ``sink" and move across the wind streamlines. These
clouds remain close to the disk surface, similar to the low-lying cloud
trajectory in Figure \ref{fig:cartoon}, with their motions closely mirroring
the disk Keplerian velocity field \citep{Kartje99}. The two families of cloud
trajectories arise from the difference in dynamics for different cloud column
densities. The ratio of ram pressure and gravitational forces on a cloud is
$F\sub{ram}/F\sub{grav} \sim \Ncrit/\Nc$. Ram pressure is the stronger force
when $\Nc \la \Ncrit$ while gravity dominates when $\Nc \ga \Ncrit$.

The analysis here mirrors the one in Section 5.1 of \cite{Kartje99} with minor
differences. Assuming roughly spherical clouds, Kartje et al introduced a
nominal mass $M\sub{c,max}$ associated with radius \Rc\ (their eq.\ 24), which
is related to \Ncrit\ via $M\sub{c,max} \sim \Ncrit R\sub{c}^2$. Then for a
cloud with mass \Mc\ and that radius, $F\sub{ram}/F\sub{grav} \sim
M\sub{c,max}/\Mc$.  But the column of such a cloud is $\Nc = \Mc/\pi
R\sub{c}^2$; thus, the results are the same since $M\sub{c,max}/\Mc \sim
\Ncrit/\Nc$. The advantage of the current formulation is in demonstrating
explicitly the complete separation between the properties of the wind (\Ncrit)
and the cloud (\Nc), and that the only relevant property of the cloud is its
column density; neither mass nor size matters separately.
}

\subsubsection{The Critical Column Density}
\label{sec:Ncrit}

If $\Mdot_{acc}$ is the accretion rate into the BLR then a fraction $\gamma$ is
carried away by the wind while the fraction $1 - \gamma$ reaches the black
hole, where it is converted to luminosity with the radiative efficiency $\eta$.
That is, $\Mw = \gamma\Mdot_{acc}$ and $L_{\rm bol} = \eta(1 -
\gamma)\Mdot_{acc}c^2$ so that $L_{\rm bol} = \eta(\gamma^{-1} - 1)\,\Mw c^2$.
The fraction $\gamma$, obtained from considerations of angular momentum
conservation for disk outflows, is \about\ 0.1--0.25 \citep{Emmering92,
Pelletier92}. {We adopt the single value $\gamma$ = 0.25 so that $L_{\rm bol}
\approx 3\eta\Mw c^2$ and get from eqs.\ \ref{eq:Ncrit1} and \ref{eq:Rd}
\eq{\label{eq:Ncrit2}
   \Ncrit(\Rd) \approx 4.3\x\E{22}{1\over\eta_{-2}}\,
            \left(\frac{L_{45}}{M_7^{2/3}}\right)^{3/4}\ \cs,
}
where $\eta_{-2} = \eta/0.01$. From this value on the BLR outer radius, \Ncrit\
increases inwards as $r^{-1/2}$ to a maximum of \hbox{\about\,10\x\Ncrit(\Rd)}
on the BLR inner boundary. The behavior of \Ncrit\ is dominated by the
variation of $\Lbol/M^{2/3}$, a quantity that spans more than 10 orders of
magnitude across the combined set of AGNs studied in
\S\ref{sec:Intermediates}.}

\subsection{Emission Characteristics}
\label{sec:F_BLR}

The specific luminosity of a broad line can be written as $\LBL\phi(\v)$, where
\LBL\ is the overall line luminosity and $\phi(\v)$ is the line profile as a
function of velocity ($\int\!\phi d\v = 1$). We discuss separately each
component.

\subsubsection{The ratio $\LBL/\Lbol$}
\label{sec:CBL}

Two factors control $\LBL/\Lbol$, the fraction of the AGN luminosity converted
into emission in a particular line. The first is the fraction of continuum
radiation intercepted by all clouds. It depends on the geometry of the BLR
cloud distribution and on the cloud optical depths, and can be considered the
BLR geometrical covering factor. The second is the efficiency with which an
individual cloud reprocesses the absorbed continuum into emission in the
particular line. Given a cloud density, column density and chemical
composition, this reprocessing is determined by the continuum spectral shape
and the magnitude of its local flux, $L_{\rm bol}/4\pi r^2 \propto (\Rd/r)^2$
(eq.\ \ref{eq:Rd}). All clouds with similar internal properties located at the
same scaled radial distance $r/\Rd$, i.e., same fraction of the BLR radial
size, will produce the same line spectrum, whatever the luminosity. Therefore
the line conversion fraction $\CBL = \LBL/\Lbol$ does not contain any explicit
dependence on luminosity; it depends only on the continuum spectral shape, the
geometry of the cloud distribution and the internal properties of individual
clouds.

\subsubsection{Line Profiles}
\label{sec:profiles}

The line profile $\phi(\v)$ is determined by the kinematics and geometry of the
cloud distribution. Detailed model calculations of broad-line profiles were
reported for two different approaches. \cite{Emmering92} analyzed line emission
from an ensemble of clouds moving along the streamlines of a hydromagnetic disk
wind in an attempt to reproduce typical properties of observed AGN broad lines.
One of the major challenges is the line shapes which show velocity widths in
excess of 10,000 \kms\ together with central cusps with widths of only \about
300 \kms. Since such sharp cusps are seen in high-ionization permitted metal
lines, e.g., {\sc C\,iv} $\lambda$1549, they cannot be solely attributed to the
contribution of a narrow component. With suitable choice of parameters,
Emmering et al were able to reproduce the observed profile asymmetries and
sharply cusped peaks when the emission was dominated by clouds at high
altitudes; emission from low-altitude clouds tended to produce double-peaked
profiles instead. \cite{Bottorff97} performed similar calculations with some
added details and applied them to the well studied Seyfert\,1 galaxy NGC 5548,
successfully reproducing all observations including line profiles and
reverberation measurements. {Since these models employed clouds moving with the
wind, they would be produced by the cloud equation of motion (eq.\
\ref{eq:EOM1}) when the gravitational force is omitted. We denote the profiles
produced from such motions \Fw.}

The other approach, motivated by the double-peaked profiles observed in some
AGNs, invoked emission from the Keplerian accretion disk \citep{Chen89,
Eracleous03, Eracleous09}. These models successfully explain many details of
double-peaked AGN observations and also show that the line profiles can turn
into single-peaked shapes by the presence of a disk wind with suitable
properties. {Purely Keplerian trajectories would be produced by the cloud
equation of motion (eq.\ \ref{eq:EOM1}) when the ram pressure force is omitted.
We denote the line profiles produced by such Keplerian motions \Fk.}

\subsubsection{Composite BLR Emission}

As shown in \S\ref{sec:trajectories}, clouds with subcritical columns, $\Nc <
\Ncrit$, move along the streamlines of disk winds, therefore the profile of
line emission from an ensemble of such clouds is \Fw. If the line conversion
factor (the fraction of ionizing radiation which is reprocessed and emitted by
the BLR; \S\ref{sec:CBL}) of the cloud distribution is \Cw, the specific line
luminosity is $L_{\rm bol}\Cw\Fw$. Similarly, the motions of supercritical
clouds, $\Nc > \Ncrit$, trace the disk Keplerian rotation, therefore we can use
\Fk\ to describe the line profiles they generate. With a line conversion factor
\Ck, the specific line luminosity for an ensemble of supercritical clouds is
$L_{\rm bol}\Ck\Fk$. As noted above, the fraction of the continuum converted to
broad lines is smaller for low-lying trajectories than for the high-altitude
wind trajectories, therefore \Ck\ is significantly smaller than \Cw.

In the most general case, the BLR will contain a mix of both subcritical and
supercritical clouds. Denote by $f$ the fractional contribution of subcritical
clouds to the overall line emission, then
\eq{\label{eq:BLR}
   \frac{\LBL}{\Lbol}\phi(\v) = f\,\Cw\Fw + (1 - f)\,\Ck\Fk.
}
The mixture introduces a new fundamental element to broad-line emission. As
discussed above, with a fixed continuum spectral shape, cloud distribution and
individual cloud properties, both the profile $\phi$ and the line conversion
fraction \CBL\ are independent of luminosity. But lowering the luminosity while
keeping all other parameters fixed will lower the fraction $f$ because \Ncrit\
decreases with $L_{\rm bol}$ (eq.\ \ref{eq:Ncrit2}). Therefore every cloud with
a given column \Nc\ becomes supercritical once $L_{\rm bol}$ has fallen below
the value that yields $\Ncrit = \Nc$. With clouds shifting from the sub- to
super-critical category, the cloud distribution is changing even if all the
underlying BLR properties such as individual cloud parameters, wind
streamlines, etc., do not change. As a result, the observed line luminosity and
profile (a combination of the elevated-wind and low-disk components) are
changing with $L_{\rm bol}$.
\footnote{Integrating eq.\ \ref{eq:BLR} over \v\ yields $\CBL = f\Cw + (1 -
f)\Ck$, and the overall profile is $\phi(\v) = (\Cw/\CBL)f\Fw + (\Ck/\CBL)(1 -
f)\Fk$}

\subsubsection{Broad-line Spectral Evolution}

Column densities of BLR clouds can be expected to cover a large range, but only
those exceeding a minimal \Nmin\ \about\ 5\x\E{21} \cs\ will contribute to the
observed broad-line spectrum \citep[e.g.][]{Netzer90}. An upper bound on \Nc\
can be deduced from time variability of X-ray observations, caused by the
passage of obscuring clouds across the line of sight. \cite{Risaliti02} find
obscuring cloud columns in the range \about\E{22}--\E{23} \cs, although an
exceptional case of a Compton thick cloud ($\Nc > \E{24}\,\cs$) has also been
recorded \citep{Risaliti07}. Whatever the maximal column, it is reasonable to
assume that clouds with lower columns are more abundant, namely, the cloud
distribution rises toward lower \Nc. When the column density of a cloud at
distance $r$ from the black hole is in the range
\eq{\label{eq:subcrit}
   \Nmin \simeq 5\x\E{21}\,\cs \la \Nc \la \Ncrit(r),
}
the cloud is subcritical. The lower bound, \Nmin, is universal, set by atomic
physics, while the upper bound, $\Ncrit(r)$, varies across the BLR as
$r^{-1/2}$ (eq.\ \ref{eq:Ncrit1}), in addition to its overall variation with
the AGN luminosity and black hole mass (eq.\ \ref{eq:Ncrit2}). A declining
luminosity pushes the upper bound downward toward the lower one, shrinking the
range of subcritical columns. As a result, clouds that were subcritical at the
higher luminosity can become supercritical, their trajectories shifting away
from the wind streamlines into the low-lying, rotating category (Figure
\ref{fig:cartoon}). This reduces the fraction $f$ and shifts the BLR emission
from the first term in eq.\ \ref{eq:BLR} to the second, lowering the broad-line
strength and leading to the emergence of double-peaked profiles.

{Shrinkage of the subcritical range with decreasing luminosity provides a
natural explanation for the $1.0 \to 1.2/1.5 \to 1.8/1.9$ sequence as it
continually increases the fraction of supercritical columns in the cloud
population. The SL12 sample shows that in spite of the huge range of
$\Lbol/M^{2/3}$, observable changes in broad-line emission arise from
relatively small variations in this variable: The separation between the mean
values of the 1.0 and 1.5 sources in that sample is less than 40\%, with an
additional decrease of just under 30\% separating the 1.8s from the 1.5s
(Figure \ref{fig:SL_histograms}). This behavior indicates that the fractional
contributions of sub- and supercritical clouds to the broad-line emission are
roughly comparable in type 1 AGNs. Indeed, the mean value of
$L_{45}/M^{2/3}_{-7}$ for QSO and Seyfert 1 galaxies is 0.18 (see Figure
\ref{fig:histograms}), and $\eta$ is typically \hbox{\about 5--10\%} in these
sources \citep{Martinez-Sansigre11}. With $\eta_{-2} = 5$, the value of \Ncrit\
varies from \about\,2\x\E{22} \cs\ on the BLR inner boundary to a minimum of
2\x\E{21} \cs\ on its outer edge. Since the distribution of cloud column
densities can be expected to rise toward its lower end, subcritical columns can
dominate the emission and lead to standard type 1 spectra, with supercritical
columns still comprising a significant fraction. This explains why relatively
small changes in $\Lbol/M^{2/3}$ from its particular value in type 1 AGNs
suffice to tip the balance between the two families of trajectories, triggering
the change in spectral type.
}

At the lower luminosities of the EH09 sample, supercritical columns dominate
decisively. In type 1.8/1.9 sources, the histograms in Figure
\ref{fig:histograms} show that the mean value of $L_{\rm bol}/M^{2/3}$ is three
orders of magnitudes lower than for type 1.0. If $\eta$ were the same for both
classes, \Ncrit\ for type 1.8/1.9 would be only \about\,1\x\E{20}\,\cs\ on the
BLR inner boundary and all clouds would be supercritical everywhere. However,
in all likelihood the accretion in these sources has a low radiative
efficiency, with $\eta \la \E{-3}$ (see \citealt{Ho08} and references therein),
so \Ncrit\ can be expected to be comparable to \Nmin\ on the inner boundary
(eq.\ \ref{eq:Ncrit2}). This would still make the cloud population of type
1.8/1.9 dominated by supercritical column densities, resulting in weaker
broad-line emission together with double-peaked profiles when viewed from
suitable angles. Further decrease of the accretion rate triggers the mass
conservation limit found in \cite{ElitzurHo09}, broad-line emission disappears
and the AGN becomes a true type 2.

\section{Summary and Discussion}

Intermediate-type AGNs have been generally dismissed as a minor detail of
unification. Instead, the data now show that as the accretion rate decreases,
broad-line emission progressively declines from its high intensity in quasars
and Seyfert 1 galaxies to the lower relative strength of intermediate-type
AGNs, disappearing altogether at sufficiently low luminosities. As summarized
in Figure \ref{fig:CFplot}, a decreasing accretion rate is accompanied by a
decrease in the fraction of bolometric luminosity converted into broad-line
emission. This systematic behavior is at odds with the fine-tuning required by
explanations of the intermediate types that were based on partial obscuration
or transient radiative transfer effects \citep[e.g.,][]{Korista04}. It also
cannot be explained by a change in the continuum spectral shape. Although the
disappearance of the big blue bump causes significant variation of the spectral
energy distribution from high- to low-luminosity AGNs (see, e.g., Ho 2008,
Figure 7), an explanation based on such changes could not apply to the SL12
data set. Thanks to a sizeable number of sources, this sample provides decisive
evidence for transitions to intermediate spectral classes at fairly high
luminosities---the mean luminosity is 3\x\E{44} \erg\ for the sample type 1.0
objects, 1\x\E{44} \erg\ for the 1.5s and 4\x\E{43} \erg\ for the 1.8s. There
are no significant changes in continuum spectral shape among such luminosities;
indeed, such changes do not show up across the entire SL12 data set, whose
luminosity range is \Lbol\ = 3\x\E{42}--2\x\E{46} \erg\ (Stern \& Laor 2012a).
The spectral sequence $1 \to 1.2/1.5 \to 1.8/1.9 \to 2$ is a true evolutionary
sequence, reflecting evolution of the BLR structure with decreasing accretion
rate onto the central black hole. The emergence of double-peaked profiles along
the sequence is additional support for such an inherent structural change.

According to standard unification, the torus obscuration and the observer
location are the only factors in determining the spectral class of any given
AGN. There is no a priori expectation for the broad line spectrum to correlate
with $L_{\rm bol}/M^{2/3}$. Yet the spectral classes of broad-line emitters
cluster around different values of $L_{\rm bol}/M^{2/3}$ and this variable
controls the evolutionary sequence of type 1 $\to$ 1.2/1.5 $\to$ 1.8/1.9 $\to$
2. Additionally, double-peaked profiles emerge along this sequence. Both trends
find a natural explanation in the disk-wind scenario for the BLR: as the
luminosity decreases, more and more clouds shift from high-altitude
trajectories along the wind streamlines to low-altitude motion that follows
closely the disk rotation (Figure \ref{fig:cartoon}). This reduces the fraction
of the continuum intercepted by the BLR---hence the reduction in line
strengths---and leads to the emergence of double-peaked emission. The
transition between the two families of trajectories is controlled by a
wind-determined critical column \Ncrit\ (eqs.\ \ref{eq:Ncrit1} and
\ref{eq:Ncrit2}); this accounts for the $L_{\rm bol}/M^{2/3}$ dependence. This
transition happens even if neither cloud properties nor wind structure change
with luminosity. It occurs under a broad range of conditions, as it arises
simply from  the competition between the black hole gravity and the wind ram
pressure, irrespective of the detailed mechanisms that drive and control the
outflow. Furthermore, this mechanism properly explains the values of $L_{\rm
bol}/M^{2/3}$ at the transitions from one spectral class to the next in spite
of the huge variation range of this variable that spans more than 10 orders of
magnitude.  The quantity that sets the scale for these transitions is \Nmin\
\about\ 5\x\E{21}\,\cs, the minimal column density for clouds to partake in
broad line emission. \Nmin\ is determined purely by atomic constants, entirely
independent of any considerations that involve the BLR detailed structure. This
fundamental quantity sets the proper scale for the transitions between spectral
classes without any free parameters or fine tuning, another success of the
disk-wind scenario.

It is important to note that exceptions exist: not every double-peaked object
has a low Eddington ratio and not every intermediate-type AGN has a low
luminosity. Such cases are to be expected since the BLR structure is anything
but simple. Still, the trends are quite clear and fit well into the outflow
scenario. While the variable $L_{\rm bol}/M^{2/3}$, which is close to the
Eddington ratio, emerges naturally, it controls only one aspect of the disk
outflow; other dynamical effects may introduce additional dependencies. For
example, \cite{Nicastro00} considered the interplay between radiation pressure
and gas pressure on the outflow and found a dependence on $L_{\rm
bol}/M^{7/8}$, practically indistinguishable from $L_{\rm bol}/M^{2/3}$.
Similarly, the growing importance of radiatively inefficient accretion flows at
lower Eddington ratios could also play a role in the BLR evolution, especially
in governing the details of the transition between various spectral classes
\citep{Ho08, Trump11}.  Moreover, while the bulk of the data shows diminishing
broad-line emission with accretion rate, two recent discoveries buck the trend.
\cite{Ho12} and \cite{Miniutti13} have found AGNs with very low black hole
masses (\E5--\E6\,\Mo) and very high Eddington ratios, but no broad lines. In
both cases, the absence of broad lines is highly significant. Because of their
high Eddington ratios, both objects are above the EH09 boundary (Figure
\ref{fig:L-M-Edd}), an absolute limit that follows from mass conservation. This
limit implies that there should be only true type 2 objects below the EH09
boundary, but it does not preclude true type 2 AGNs above it. Still, these
objects do not fit the general observed pattern and are not explained by our
scenario. They require another explanation, and some possibilities are
discussed in \cite{Miniutti13}.

Model calculations employing clouds moving along disk-wind streamlines
successfully explain details of broad-line profiles, including their
difficult-to-explain sharply cusped peaks \citep{Emmering92}. Model
calculations of disk emission successfully explain the double-peaked profiles
observed in some AGNs \citep{Eracleous09}. The first class of models
corresponds to $f = 1$ in Eq.\ \ref{eq:BLR}, the second to $f = 0$; therefore
our model accrues all the benefits of both earlier models at the two proper
ends of the accretion rate spectrum. A full calculation of the combined
profiles and line strengths with accretion rate is a complex task. Other broad
emission lines may also deviate slightly from the behavior observed for \Ha\
since the supercritical clouds may receive insufficient ionizing flux to
support the higher-ionization broad lines \citep[e.g.,][]{Proga00, Proga04}.
This arises from the combined effect of a decrease of the ionization parameter
at low altitudes because of the continuum anisotropy and the change in the
ionizing continuum spectral shape at low luminosities. We plan on performing
such detailed calculations and will report their results in future
publications.

\subsection{Observational Challenges}
\label{sec:Observations}

Observations of intermediate-type AGNs present unique challenges. Based on line
spectra alone, NGC 1068 would be classified as type 1.8, NGC 4388 as 1.9. Yet
both are Seyfert 2 galaxies with  obscured nuclei (Compton-thick obscuration in
the case of NGC 1068) whose weak broad lines reflect scattered light, as
determined by polarization measurements \citep{Ski85, Young96c}. In an inherent
ambiguity, type 1.8/1.9 spectra can arise from either the directly-arriving
weak broad-line emission of an unobscured nucleus, intrinsically a type 1.8/1.9
AGN, or indirectly through the scattering of strong broad lines from an
obscured nucleus. The latter is an obscured type 2 AGN appearing as a spurious
1.8/1.9 object because of a hidden central engine whose intrinsic broad line
strength is that of a type 1 source. Spectral information alone is insufficient
to resolve the ambiguity of these two very different situations. A decisive
determination of the true nature of a type 1.8/1.9 spectrum requires also a
measurement of the obscuration to the nucleus, ideally supplemented by a
measurement of the broad-line polarization.

{Since \Ncrit\ varies as $r^{-1/2}$ (eq.\ \ref{eq:Ncrit1}), the transition from
high-altitude wind trajectories to low-altitude disk-like rotation when \Lbol\
is decreasing should occur first at larger radii. Lines emitted further out in
the BLR should thus be the first to display reduced relative strengths and
double-peaked emission profiles at lower luminosities. This provides a testable
prediction of the mechanism proposed here.
}

Much of the support for the data analysis presented here comes from the SL12
sample, which was culled out of the SDSS database. For the most part, these
observations lacked the angular resolution to separate the AGN from the host
galaxy and could not identify AGNs with luminosities below \about \E{42}\,\erg,
where the bulk of the intermediate-type population is expected. But what the
SL12 sample lacked in luminosity coverage it made up in sheer size, showing the
power of large samples to reliably identify trends in the data. As can be seen
from Figure \ref{fig:L-M-Edd}, the crucial low-luminosity end where most
intermediate-type AGNs can be expected to reside and most of the BLR evolution
to occur is covered only sparsely in comparison. Future observations with
higher angular resolution are needed to fill this gap and provide more complete
coverage of the full evolution of the AGN broad line region.

\smallskip

\paragraph*{Acknowledgements:}
We thank Jonathan Stern and Lisa Winter for providing us with their data.
Special thanks to Ari Laor and Jonathan Stern for their help and useful
comments, and to Jorge Pineda for help with the KS test. M.E. acknowledges the
award of an NPP Senior Fellowship from ORAU, which supported a sabbatical leave
at JPL/Caltech where much of this work was completed.  The work of L.C.H. is
supported by the Kavli Foundation, Peking University, the Chinese Academy of
Sciences, and Carnegie Institution for Science. J.R.T. is supported by NASA
through Hubble Fellowship grant HST-HF-51330.01 awarded by the Space Telescope
Science Institute, which is operated by the Association of Universities for
Research in Astronomy, Inc., for NASA, under contract NAS 5-26555.


\label{lastpage}

\end{document}